\documentclass[reprint,prd]{revtex4-2}
\usepackage{hyperref}
\hypersetup{colorlinks,allcolors=black}
 \usepackage{graphicx}
 \usepackage{float}
 \usepackage{amsmath}
 \usepackage{amssymb}
 \usepackage[utf8]{inputenc}
 \usepackage{natbib}
 \usepackage{slashed}
\begin{document}
\title{The Natural TeV Cutoff of the Higgs Field from the Multiplicative Lagrangian}
	\author{Suppanat Supanyo$^{1,2}$}
 	\author{Monsit Tanasittikosol$^{1,2}$}
 		\author{Sikarin Yoo-Kong$^3$}
 		\affiliation{$^1$Theoretical and Computational Physics (TCP) Group, Department of Physics, Faculty of Science, King Mongkut's University of Technology Thonburi, Bangkok 10140, Thailand}
 		\affiliation{$^2$Theoretical and Computational Science Centre (TaCS), Faculty of Science, King Mongkut's University of Technology Thonburi, Bangkok 10140, Thailand}
 	\affiliation{$^3$The institute for Fundamental Study (IF), Naresuan University, Phitsanulok 65000, Thailand}
 	\begin{abstract}
 	The various types of the non-standard Lagrangian can be added to the standard Lagrangian with the invariant of the equation of motion in the low energy limit. In this paper, we construct the multiplicative Lagrangian of a complex scalar field giving the approximated Klein-Gordon equation from the inverse problem of the calculus of variation.  Then, this multiplicative Lagrangian with arbitrary high cutoff is applied to the toy model of the Higgs mechanism in U(1)-gauge symmetry in order to study the simple effects in the Higgs physics. We show that, after spontaneous symmetry breaking happens, the Higgs vev is free from the Fermi-coupling constant and the Higgs field  gets the natural cutoff in TeV scale. 
 	The other relevant coupling constants, the UV-sensitivity of Higgs mass due to the loop correction, some applications on the strong CP problem as well as anomalous small fermion mass, and the cosmological constant problem are also discussed.
\end{abstract}
\maketitle

\section{Introduction}
The renormalizability of the Standard Model (SM) is the successful description to unify the three fundamental forces: electromagnetic force, weak force, and strong force \cite{originOfHiggs1,originOfHiggs2,originOfHiggs3,oringPfHiggs4,Electroweak1,Electroweak2,Electroweak3}. From the unitarity, stability, and triviality \cite{Holthausen:2011aa,Bezrukov:2012sa,Masina:2012tz,Degrassi:2012ry}, this model can be valid up to the Planck scale. The ultra-violet cutoff (UV-cutoff) $\Lambda_{\rm UV}$ from the loop corrections can possibly set into arbitrary high energy scale and can be absorbed into the counterterms. However, many phenomena cannot be included in the SM prediction such as the quantum gravity, the non-vanishing neutrino mass, the strong CP-problem  et al. \cite{Robertson:2015owa,Bilenky:2016pep,PhysRevLett.44.912,PhysRevLett.43.1566}.  Hence, it is reasonable to interpret the SM as the effective theory below the unknown energy scale.
For this reason, the concept of naturalness comes to play an essential role for specifying the mass scale of new physics  \cite{Susskind:1978ms,HeirarchyProblem,Williams:2015gxa,dijkstra2019naturalness,rosaler2019naturalness,vanKolck:2020plz}.  The successful cases are the prediction of the mass scale of the charm quark and the rho meson before they were discovered.  The warning sign from the existence of the new physics in the effective field theory is that the unitarity of the scattering amplitude is violated by some particular scale of the parameter $\Lambda_{\rm UV}$. Here, the notion naturalness can be defined as the autonomy of the physical scale (AOS) of the effective theory \cite{Williams:2015gxa}. Namely, the parameters in the effective field theory should not sensitive to the physics in the UV-scale. For example, in the case of effective theory of fermion, the radiative correction to mass of the fermion ($M$) is proportional to the mass itself: $M\log(\Lambda_{\rm UV}^2/M^2)$. The UV-cutoff in the logarithmic function leads to a small quantum correction although the $\Lambda_{\rm}$ is higher than $M$ many orders of magnitude and hence, the fermion mass is not sensitive to the physics at the UV-scale. Therefore, the fermion is natural in this sense. On the other hand,  in the case of the scalar field, the radiative correction to scalar mass ($m$) is proportional to the quadratic UV-cutoff $\Lambda_{\rm UV}^2$. The mass parameter of scalar field is sensitive to the physics at UV-scale. This causes that the scalar field theory is unnatural and might not be the elementary \cite{Susskind:1978ms}. To restore the naturalness of the scalar field, the bare mass and the physical mass are required in the same order magnitude as the UV-cutoff $m_{\rm physical}\sim m\sim \Lambda_{\rm UV}$ \cite{Susskind:1978ms}. Here, the naturalness problem of scalar field is in an unpleasant situation after the elementary scalar particle: Higgs boson with a mass $125$ GeV \cite{HiggsMass125,HiggsMass1252} was discovered at LHC.  The unnatural situation appears in the Higgs sector in the SM. The radiative correction to the bare Higgs mass parameter $\mu$ is proportional to the UV-cutoff $\Lambda_{\rm UV}$
\begin{align}\label{1}
    M_{Higgs, \text{physical}}^2=2\mu^2+c\Lambda^2_{\rm UV}+...
\end{align}
where $ M_{Higgs, \text{physical}}\approx 125$ GeV is an observed Higgs mass, $c$ contains the linear combination of the square mass of the particle in the SM, and the rest contains the logarithmic term and a series of the inverse UV-cutoff. If the SM is assumed to be valid below the quantum gravity scale, it is reasonable to take the UV-cutoff to be the Planck Mass.  The fine-tuning problem around 36-orders of decimal is required to get the small physical mass $(10^2)^2\approx(10^{36})-(10^{36})$. Therefore, the mass parameter of scalar boson is sensitive to the UV-physics.  To avoid the fine-tuning problem, artificially lowering cutoff energy  down to the early TeV scale is a simple solution. This situation leads to a question: Why is the cutoff energy so low comparing to the Planck scale, if the quantum gravity is a single extended part of the SM ? This problem is famously known as the Hierarchy problem \cite{HeirarchyProblem,rosaler2019naturalness}. Therefore, if the notion of the naturalness is still applicable, the SM should be replaced or gets the contributions by the new physics at TeV scale. The famous scheme for searching the new physics beyond the SM from the bottom-up point of view  is the Standard Model Effective Theory. This model is assumed that there exist heavy particles, coupling to the SM particle, in arbitrary high energy scale. By integrating out these overall heavy-particles, the SM Lagrangian should consist of the higher dimension operators $\mathcal{O}_d$ with the cutoff $\Lambda$  as
\begin{align}
    \mathcal{L}=\mathcal{L}_{\rm SM}+\frac{\hat{\mathcal{O}_5}}{\Lambda}+\frac{\hat{\mathcal{O}_6}}{\Lambda^2}+... .
\end{align}
These terms make the unitarity violation of the scattering amplitude  at the energy scale $E\sim\Lambda$ then the UV-cutoff can be defined at the same scale with the parameter $\Lambda$. From the notion of naturalness, this ends up with the expectation of $\Lambda$ in a few TeV scale. However, the various observations show that the cutoff scale of the effective operators is higher than the expectation value from naturalness argument \cite{Ellis:2018gqa,ellis2021smeft}.  Furthermore, the most famous solution: supersymmetry, was proposed to protect the mass of Higgs boson from the UV-sensitivity. The quadratic UV-cutoff is canceled by the supersymmetric particles in a few TeV mass scale. The remaining is the UV-cutoff in the logarithmic function which is roughly of order unity. Therefore, the naturalness problem is elegantly solved from the supersymmetric theory. However, no sign of supersymmetric particles has been observed at the Large Hardron Collider (LHC) \cite{aad2012search,aad2014search,aad2015search,aaboud2016search}.

We focus on the contribution terms from an important theorem in the Lagrangian mechanism, known as the non-uniqueness of Lagrangian, to fulfil the notion of the naturalness. Briefly, the Lagrangian mechanics is the mathematical tool to provide the equation of motion (EOM) ranging from the point particle system to the field theory. In the standard scheme, the Lagrangian function is written as the additive form of the kinetic $T$ and potential $V$ function, $\mathcal{L}_{\rm Standard}=T-V$. However, the inverse problem of the calculus of variation shows that we could possibly have various fascinating forms of the Lagrangian in order to explain the evolution of the system \cite{mizel1995nonuniqueness,degasperis2001newton, ELNABULSI2015120, musielak2008standard}. Many non-standard forms of Lagrangian, which are frequently seen in the cosmology  \cite{PhysRevLett.85.4438, PhysRevD.63.103510, PhysRevD.103.043518}, are proposed to explain the nature. For example, the k-essence and Dirac-Born-Infeld Lagrangians can provide the standard relativistic motion of field, the Klein-Gordon equation, added with the extra terms from the high energy effect. In general, we expect that every non-standard Lagrangian of scalar field could be reduced to the Klein-Gordon equation in low energy limit.  In this work, we hypothesise that the Lagrangian of the Higgs field may not be in either the pure standard-additive form or pure non-standard form. The Lagrangian might be written in the linear combination of many possible forms
\begin{align}\label{xxx}
    \mathcal{L}_{\phi}=\alpha_0\mathcal{L}_\text{Standard}+\sum_{i}\alpha_i\mathcal{L}_{i,\text{Non-standard}},
\end{align}
where $\alpha_i$ is a dimensionless coefficient. The EOM from Lagrangian \eqref{xxx} will be approximately the Klein-Gordon equation 
\begin{align}
    \text{EOM}(\mathcal{L}_\phi)\approx \left(\partial_\mu\partial^\mu \phi+\frac{\partial V }{\partial \phi^*}\right)=0.
\end{align}
 If the interaction from the non-standard Lagrangian could not be matched with the physical phenomena or  experimental observation, we can set $\alpha_i=0$. We then consider the contribution from the multiplicative form of Lagrangian \cite{surawuttinack2016multiplicative,PhysRevD.63.103510,PhysRevD.103.043518}, constructing from the inverse problem of the calculus of variation \cite{douglas1941solution,hojman1981inverse,sarlet1982helmholtz}, in the complex scalar field model. 
Now, let us give an overview what we are going to do. We aim to show that when the contribution terms from the multiplicative Lagrangian with the cutoff at arbitrary high energy scale ($\Lambda_H$) are added to the standard Lagrangian,  the natural cutoff scale of the Higgs field ($h$) after spontaneous symmetry breaking (SSB) can be tuned down into the TeV scale
\begin{align}
    \frac{1}{\Lambda_H^2}\mathcal{O}_6(\phi)\to \frac{1}{\Lambda_{\rm TeV}^2}\mathcal{O}_6(h),
\end{align}
where $\Lambda_H$ may possibly be set to be the Planck scale ($M_p=2.44 \times 10^{18}$ GeV). This then could reconcile the size of the radiative correction automatically.  We would like to emphasize that the result in this paper will not be used to solve the UV-sensitivity of the Higgs mass from the heavy-particles beyond SM. But we rather show that the cutoff of Higgs field can be possibly and naturally low.

This paper is organized as follows. In Sec-2, we rewrite the Lagrangian of the complex scalar field model in the non-standard form. The non-uniqueness principle of Lagrangian and the inverse calculus variation are applied to construct the Lagrangian producing the Klein-Gordon equation in arbitrary potential.  In Sec-3,  we do not work on the full SM Lagrangian to avoid abundant terms which do not answer our research problem. We apply the contribution from the multiplicative Lagrangian to study the toy-model of Higgs mechanism in U(1) gauge symmetry breaking and analogize some parameters to the electroweak theory.  The cutoff energy in the Higgs sector both before and after SSB is analysed. In Sec-4, we discuss on the coupling constants of the relevant operator, some applications, and the cosmological constant term.
\section{The multiplicative Lagrangian of complex scalar field}

We consider the multiplicative form of Lagrangian for the complex scalar field model given by
\begin{align}\label{L1}
    \mathcal{L}=F(\partial_\mu \phi^*\partial^\mu\phi)f(\phi^*\phi),
\end{align}
where $\phi$ is the complex scalar field, $\partial_\mu \phi^*\partial^\mu\phi$ is the kinetic energy of the complex scalar field. The functions $F(\partial_\mu \phi^*\partial^\mu\phi)$ and $f(\phi^*\phi)$ are unknown and to be determined. To apply this Lagrangian as the complex scalar field in the relativistic theory, we require that
the EOM from Eq.~\eqref{L1} is the approximated Klein-Gordon equation in the potential $V$ as
\begin{align}\label{EOM1}
    \partial_\mu\partial^\mu \phi+\frac{\partial V }{\partial \phi^*}\approx 0,
\end{align}
where $V= V (\phi^*\phi)$ is the potential of the complex scalar field with the global U(1) symmetry. We follow the scheme from \cite{surawuttinack2016multiplicative} to find the expressions of $F$ and $f$ using separation variable method. The functions $F$ and $f$ can be solved by the inverse calculus of variation from the Euler-Lagrange equation
\begin{align}\label{EL}
    0=\frac{\partial \mathcal{L}}{\partial\phi^*}-\partial_\mu \frac{\partial \mathcal{L}}{\partial\partial_\mu\phi^*}.
\end{align}
Substituting Eq.~\eqref{L1} into Eq.~\eqref{EL} and applying the chain rule to reorganize the spacetime derivative, $\partial_\mu$, we have
\begin{align}\label{EOM3}
    0=&F \frac{\partial f}{\partial\phi^*}-\partial_\mu\partial^\mu\phi f\frac{\partial F}{\partial X}-X \frac{\partial f}{\partial \phi^*}\frac{\partial F}{\partial X}-\partial_\mu\phi\partial^\mu\phi\frac{\partial f}{\partial\phi}\frac{\partial F}{\partial X}\nonumber
    \\
    &-\partial_\mu X\partial^\mu \phi f \frac{\partial^2 F}{\partial X^2},
\end{align}
where we have defined $X=\partial_\mu \phi^*\partial^\mu\phi$.
The $\partial_\mu\partial^\mu\phi$ can be reorganized into the function of $\phi$ by the EOM. However, the problematic terms $\partial_\mu\phi\partial^\mu\phi$ and $\partial_\mu X\partial^\mu \phi$ lead to the failure of the separation variable method (SVM). Thus, it seems to suggest that if the SVM is still applicable, this particular term must vanish
\begin{align}\label{condition1}
    \frac{\partial^2 F}{\partial X^2}=0,
\end{align}
and the fourth term can be interpreted as the extra interaction from the non-standard Lagrangian. Now, we ignore the extra interaction for a moment and we expect to obtain the Klein-Gordon equation from the first three terms in Eq.~\eqref{EOM3}. From the SVM, we have
\begin{align}
    &\frac{d F}{dX}=\frac{1}{\epsilon\Lambda^4} \left(F-X\frac{d F}{dX}\right)\label{F1},
    \\
    &\frac{d f}{d\phi^*}=-\frac{1}{\epsilon\Lambda^4} \frac{\partial V}{\partial\phi^*}f\label{f1},
\end{align}
where $\Lambda^4$ is an arbitrary positive constant and $\epsilon=\pm 1$.
By solving Eq.~\eqref{F1} and Eq.~\eqref{f1}, the functions $F$ and $f$ are expressed as
\begin{align}
    &F(X)=\epsilon\Lambda^4+X,
    \\
    &f(\phi^*\phi)=e^{-\frac{V(\phi^*\phi)}{\epsilon\Lambda^4}}.
\end{align}
Therefore, the multiplicative form of the Lagrangian of the complex scalar field is given by
\begin{align}\label{MLc}
    \mathcal{L}_{\Lambda}=\left(\epsilon\Lambda^4+\partial_\mu\phi^*\partial^\mu\phi\right)e^{-\frac{V(\phi^*\phi)}{\epsilon\Lambda^4}}.
\end{align}
Here, from the dimensional analysis, $\Lambda$ is a constant with the mass dimension $[\Lambda]=1$. The $\Lambda$ can be interpreted in two different ways. First, let's us consider the $\Lambda$ in the exponential function. The $\Lambda$ can be interpreted as the effective cutoff energy of the complex scalar field, $\phi$. Second, by considering the $\Lambda$ in the bracket, the $\Lambda$ can be interpreted as the classical value of the vacuum energy density when we take $\phi\to 0$.  

One can show that the EOM of this Lagrangian \eqref{MLc} is the Klein-Gordon equation, with the extra interaction, multiplying with the exponential function 
\begin{align}
    \left(\partial_\mu\partial^\mu\phi+\frac{\partial V}{\partial\phi^*}+\frac{\partial_\mu\phi\partial^\mu\phi}{\epsilon\Lambda^4}\frac{\partial V}{\partial \phi}\right)e^{-\frac{V}{\Lambda^4}}=0.
\end{align}
In the limit $\Lambda^4\gg \partial_\mu\phi\partial^\mu\phi$, the Klein Gordon equation is recovered
\begin{align}
    \left(\partial_\mu\partial^\mu\phi+\frac{\partial V}{\partial\phi^*}\right)e^{-\frac{V}{\Lambda^4}}\approx 0.
\end{align}
It is obvious that this Lagrangian can be another choice to explain the complex scalar field in the relativistic theory. As a consequence, the Lagrangian of the complex scalar field can be written in more general form allowed by the non-uniqueness theorem of Lagrangian. From our hypothesis in Eq.~\eqref{xxx}, one can consider a linear combination between the multiplicative Lagrangian and the linear Lagrangian 
\begin{align}\label{LinearCL}
    \mathcal{L}_\phi=\partial_\mu\phi^*\partial^\mu\phi-V+\alpha\left(\epsilon\Lambda^4+\partial_\mu\phi^*\partial^\mu\phi\right)e^{-\frac{V}{\epsilon\Lambda^4}}, 
\end{align}
while the EOM is still invariant in the limit $\Lambda^4\gg \partial_\mu\phi\partial^\mu\phi$. Here, the parameter $\alpha$ can be $\pm 1$. In classical level in the limit $\Lambda^4\gg \partial_\mu\phi\partial^\mu\phi$, the solution of field does not change. However, the multiplicative Lagrangian in Eq.~\eqref{LinearCL} can be interpreted and promoted as a boundary term of the UV-cutoff energy of the  scalar particle. For example, in case of standard massive free scalar field ($\alpha=0$) with $V=m^2\phi^*\phi$,
\begin{align}
    \partial_\mu\phi^*\partial^\mu\phi-m^2\phi^*\phi,
\end{align}
the highest energy of the scalar particle is not bounded in the perturbative theory and can be set into arbitrary high energy or infinity.  Here, in case of added multiplicative Lagrangian, there are the contribution terms from the dimension-6 operator  such as
\begin{align}
    \frac{m^6}{\Lambda^8}(\phi^*\phi)^3.
\end{align}
This term gives us the amplitude proportional to the energy of scalar particle (denoted by $E$) as
\begin{align}
    \mathcal{A}\sim\frac{m^6E^2}{\Lambda^8}.
\end{align}
The unitarity of amplitude violates at the energy scale $E> \Lambda^4/m^3$. Obviously, the highest energy of scalar particle cannot be infinite in the perturbative calculation. The UV-cutoff $\Lambda_{\rm UV}$ is obtained in the scale $\Lambda_{\rm UV}\sim \Lambda^4/m^3$. Therefore, mathematically, the UV-cutoff of the free scalar particle is given by the added multiplicative Lagrangian.

In the next section,  we apply this Lagrangian \eqref{LinearCL} to
study the toy model of Higgs mechanism in U(1) gauge symmetry breaking and analogize some parameters to the electroweak theory. 
\section{The toy model: Higgs mechanism with the multiplicative Lagrangian}
The potential $V$ in Eq.~\eqref{LinearCL}  is defined as the Higgs potential in the Ginzberg-Landau type 
\begin{align}\label{Higgspotential}
    V=-\mu^2\phi^*\phi+\lambda (\phi^*\phi)^2,
\end{align}
where $\phi$ is analogy to be the Higgs field, $\mu$ is the mass parameter, and $\lambda$ is dimensionless coupling constant. 
Substituting Eq.~\eqref{Higgspotential} into Eq.~\eqref{LinearCL}, one obtains
\begin{align}\label{LinearCL2}
    \mathcal{L}_{\phi}=&\partial_\mu\phi^*\partial^\mu\phi+\mu^2\phi^*\phi-\lambda (\phi^*\phi)^2\nonumber
    \\
    &+\alpha\left(\epsilon\Lambda^4+\partial_\mu\phi^*\partial^\mu\phi\right)e^{-\frac{-\mu^2\phi^*\phi+\lambda (\phi^*\phi)^2}{\epsilon\Lambda^4}}.
\end{align}
 The EOM of $\phi$ from the Lagrangian \eqref{LinearCL2} is given by
\begin{align}\label{EOMHiggs}
    &\partial_\mu\partial^\mu\phi-\mu^2 \phi+2\lambda (\phi^*\phi)\phi\nonumber
    \\
    &=\frac{\alpha}{\epsilon}\frac{-\mu^2 \phi+2\lambda (\phi^*\phi)\phi}{1+\alpha e^{-\frac{-\mu^2\phi^*\phi+\lambda(\phi^*\phi)^2}{\epsilon\Lambda^4}}}\frac{\partial_\mu\phi\partial^\mu\phi}{\Lambda^4} e^{-\frac{-\mu^2\phi^*\phi+\lambda(\phi^*\phi)^2}{\epsilon\Lambda^4}}.
\end{align}
The vacuum solution of field, which relates to the tree-level Higgs vev, can be obtained by setting $\partial_\mu\phi=0$ in the EOM \cite{Jones:2017ejm}
\begin{align}
    -\mu^2 \phi+2\lambda (\phi^*\phi)\phi=0.
\end{align}
The Higgs vev is non-zero given by
\begin{align}\label{vev}
   2\langle \phi^*\phi\rangle=v^2= \frac{\mu^2}{\lambda}.
\end{align}
Here, the non-derivative term in Eq.~ \eqref{EOMHiggs} can be interpreted as the field derivative of potential 
\begin{align}\label{VHiggss}
    \frac{\partial V_{\rm Higgs }}{\partial\phi^*}=-\mu^2 \phi+2\lambda (\phi^*\phi)\phi.
\end{align}
 We shall discuss about the shape of the full Higgs potential and the energy density of theory after we fit the parameter and obtain the UV-cutoff of Higgs field in the discussion section. Then, we assume that the Lagrangians of the other particles are in the standard form. We add the U(1) gauge field ($A_\mu$), Wely fermion ($\psi$), and the $U(1)$ covariant derivative to the Lagrangian \eqref{LinearCL} 
\begin{align}\label{Ltoy1}
    &\mathcal{L}=-\frac{1}{4} F_{\mu\nu} F^{\mu\nu}+D_\mu\phi^*D^\mu\phi+\mu^2\phi^*\phi-\lambda (\phi^*\phi)^2\nonumber
    \\
    &+\alpha\left(\epsilon\Lambda^4+D_\mu\phi^*D^\mu\phi\right)e^{-\frac{-\mu^2\phi^*\phi+\lambda (\phi^*\phi)^2}{\epsilon\Lambda^4}}\nonumber
    \\
    &+i\bar{\psi}_L \slashed{D} \psi_L+i\bar{\psi}_R\slashed{D}\psi_R-\lambda_f\left(\phi\bar{\psi}_L\psi_R+\phi^*\bar{\psi}_R\psi_L\right),
\end{align}
where $ D_\mu=\partial_\mu -ig A_\mu$, $F_{\mu\nu}=\partial_\mu A_\nu-\partial_\nu A_\mu$, and $\lambda_f$ is the Yukawa-coupling constant.  The U(1)-symmetry is in the meta-stable state. This then results in the existence of the spontaneous symmetry breaking. We expand the Higgs field around the vev as
\begin{align}
    \phi(x)=\frac{v+h(x)}{\sqrt{2}}e^{i \xi(x)},
\end{align}
where $h(x)$ and $\xi(x)$ are the fluctuation of field in the radial part and the angular part in the complex plan. We apply the unitary gauge $\xi (x)=0$ and, consequently, the Goldstone mode can be ignored. After SSB, the kinetic energy term of Higgs field is in the non-canonical form scaling by the function $G(h)$ as
\begin{align}
    \frac{1}{2}G(h) \partial_\mu h\partial^\mu h,
\end{align}
where
\begin{align}
    G(h)=1+\alpha  e^{-\frac{\lambda (h+v)^4-2 \mu ^2 (h+v)^2}{4 \epsilon \Lambda ^4 }},
\end{align}
 which contains the exponential factor. We can reorganize this term into the canonical renormalized form by redefining the field  as
\begin{align}
h\to \beta h+\alpha \beta^5\frac{  \mu
   ^2 }{6 \Lambda ^4 \epsilon }e^{\frac{\mu ^2 v ^2}{4 \Lambda ^4 \epsilon }}h^3+O\left(h^5\right),
\end{align}
where 
\begin{align}\label{beta}
    \beta=\frac{1}{\sqrt{1+\alpha  e^{\frac{\mu ^2 v ^2}{4 \Lambda ^4 \epsilon }}}}.
\end{align}
The Lagrangian in Eq.~\eqref{Ltoy1} after SSB can be expressed as
\begin{align}
    \mathcal{L}=&-\rho_{\rm Higgs}+\frac{1}{2}\partial_\mu h \partial^\mu h\nonumber-\frac{M_h^2}{2} h^2-\frac{\lambda_3}{3!}h^3-\frac{\lambda_4}{4!}h^4
    \\\nonumber
    &-\frac{4\pi}{\Lambda_5}h^5-\frac{(4\pi)^2}{\Lambda^6}h^6+\mathcal{O}(h)^7
    \\\nonumber
    &-\frac{1}{4} F_{\mu\nu} F^{\mu\nu}-\frac{1}{2} M_A^2 A_{\mu} A^{\mu}+g_{hAA}h A_{\mu}A^\mu+\mathcal{O}(h)^2
    \\\nonumber
    &+i\bar{\psi}_L \slashed{\partial} \psi_L+i\bar{\psi}_R\slashed{\partial}\psi_R-M_f\left(\bar{\psi}_L\psi_R+\bar{\psi}_R\psi_L\right)
    \\
    &-y h \left(\bar{\psi}_L\psi_R+\bar{\psi}_R\psi_L\right)+\mathcal{L}_\text{fermion-gauge}.
\end{align}
Here, we have reorganized the coefficients in front of the higher dimension operators in the form of NDA power-counting formula \cite{Manohar:1983md,Gavela:2016bzc}. The coefficients are defined as
\begin{align}
   & M_h=\sqrt{2}\mu,\quad M_f=\frac{\lambda_f v}{\sqrt{2}},\quad  M_A^2=\beta^{-2}g^2 v ^2 ,\nonumber
   \\\nonumber
   &y=\frac{\beta\lambda _f}{\sqrt{2}},\quad
    \lambda _3=\frac{6\beta \mu ^2}{v}, \quad g_{hAA}=\beta^{-1}e^2 v 
    \\\nonumber
    &\lambda _4=\frac{2\beta^4 \mu ^2 \left(\alpha  e^{\frac{\mu ^2 v^2}{4 \Lambda ^4 \epsilon }} \left(3 \Lambda ^4
   \epsilon -2 \mu ^2 v^2\right)+3 \Lambda ^4 \epsilon \right)}{\Lambda ^4 v^2 \epsilon }, 
   \\
   &\Lambda_5=\frac{16 \pi \beta^{-5}  \Lambda ^4 v \epsilon  e^{-\frac{\mu ^2 v^2}{4 \Lambda ^4 \epsilon }}}{\alpha  \mu
   ^4}\nonumber
   \\
   &\Lambda _6=\frac{24\sqrt{10}\pi\beta^{-4}\epsilon v\Lambda^4e^{-\frac{\mu ^2 v^2}{8\Lambda ^4 \epsilon }}}{\mu^2\sqrt{\alpha\left(8 (\beta^{-2}-4) \mu ^2 v ^2+57 \beta^{-2} \Lambda ^4 \epsilon\right)}},\nonumber
   \\
   & \rho_{\rm Higgs}=-\frac{\mu ^2 v^2}{4}-\epsilon\alpha  \Lambda ^4   e^{\frac{\mu ^2 v^2}{4 \Lambda ^4 \epsilon }}
   \label{parameter1}.
\end{align}
When $\Lambda$ is taken to be infinity, all parameters are reduced to the prediction in the standard theory of U(1)-Higgs mechanism. Then, we specify the Higgs vev by fitting the tree-level parameters to the Fermi coupling constant ($G_F$) as the traditional scheme. By integrating out of the gauge boson in the $ff\to ff$ scattering amplitude,  the Fermi-coupling constant can be written as,
\begin{align}\label{GF}
    \frac{G_F}{\sqrt{2}}=\frac{e^2}{8M_A^2}.
\end{align}
This expression is the same as the standard theory due to non-modification in the gauge-fermion interaction in Eq.~\eqref{Ltoy1}.
Substituting the parameter $M_A$ from Eq.~\eqref{parameter1} into Eq.~\eqref{GF}, we have
\begin{align}\label{GF2}
   G_F=\frac{1}{4 \sqrt{2} v ^2 \left(1+\alpha  e^{\frac{\mu ^2 v ^2}{4 \Lambda ^4 \epsilon }}\right)}.
\end{align}
Now, the $G_F$ cannot be applied to specify the observed value of $v$. This parameter depends on two free-parameters $v$ and $\Lambda$ while $\mu$ is analogized to be Higgs mass parameter. To match this model to the four-fermion theory, we can tune the value of $v$ and $\Lambda$ as follows
\begin{align}\label{fitLambda}
    \Lambda^4=\frac{(-1)}{\epsilon}\frac{v^2 M_h^2}{8   \log \left(-\frac{\alpha  v^2}{v^2-v_{\text{SM}}^2}\right)},
\end{align}
where we have parametrized $G_F$ in terms of $v_{\rm SM}$, which $v_{\rm SM}$ is the vev in the standard theory defined from Eq.~\eqref{GF2} at $\Lambda\to\infty$, as the following equation $v_{\rm SM}^2=1/4\sqrt{2}G_F$. Here, $v_{\rm SM}$ is analogized to the Higgs vev from the Standard Model, $v_{\rm SM}=246$ GeV \cite{EWtest}. The important remark is that $v_{\rm SM}^2$ is not the Higgs vev but it is just the parametrization of the inverse Fermi coupling constant.  According to the previous section, the definition of the parameter $\Lambda^4$ is a positive constant, see Eq.~\eqref{F1} and Eq.~\eqref{f1}, so Eq.~\eqref{fitLambda} holds if $\epsilon=-1$ is a required choice. Then, we consider a possible value of the parameter $\alpha$ in Eq.~\eqref{fitLambda}. If $v>v_{\rm SM}$, the term $- v^2/(v^2-v_{\rm SM}^2)$ is obviously negative. Thus, one requires $\alpha=-1$ to satisfy the definition of positive $\Lambda^4$. It is not difficult to see that, on the other hand, if $v<v_{\rm SM}$ the value of the parameter $\alpha$ must be $+1$. 

In the large vev limit, $\Lambda^4$ in Eq.~\eqref{fitLambda} is reorganized in the following form
\begin{align}\label{vevfit2}
    \Lambda^4\simeq \frac{v^4 M_h^2}{8 v_{\text{SM}}^2}.
\end{align}
To obtain the precise regime of the Higgs mass, the Dirac naturalness \cite{Dirac:1938mt} requires the dimensionless ratio  $v/\Lambda$ from Eq.~\eqref{vevfit2} to be of order unity
\begin{align}
    \frac{v}{\Lambda}\simeq \sqrt{\frac{v_{\rm SM}}{M_h}}\sim \mathcal{O}(1).
\end{align}
it obviously implies that the Higgs mass should be around the order of the inverse Fermi-coupling, $ v_{\rm SM}\sim10^2$ GeV,
\begin{align}\label{result1}
    M_h\sim v_{\rm SM}.
\end{align}
Eq.~\eqref{result1} compatibly matches with the observed value of  $M_h=\sqrt{2}\mu\approx 125$ GeV \cite{HiggsMass125,HiggsMass1252}.

\begin{figure}[H]
     \centering
     \includegraphics[width=0.8\linewidth]{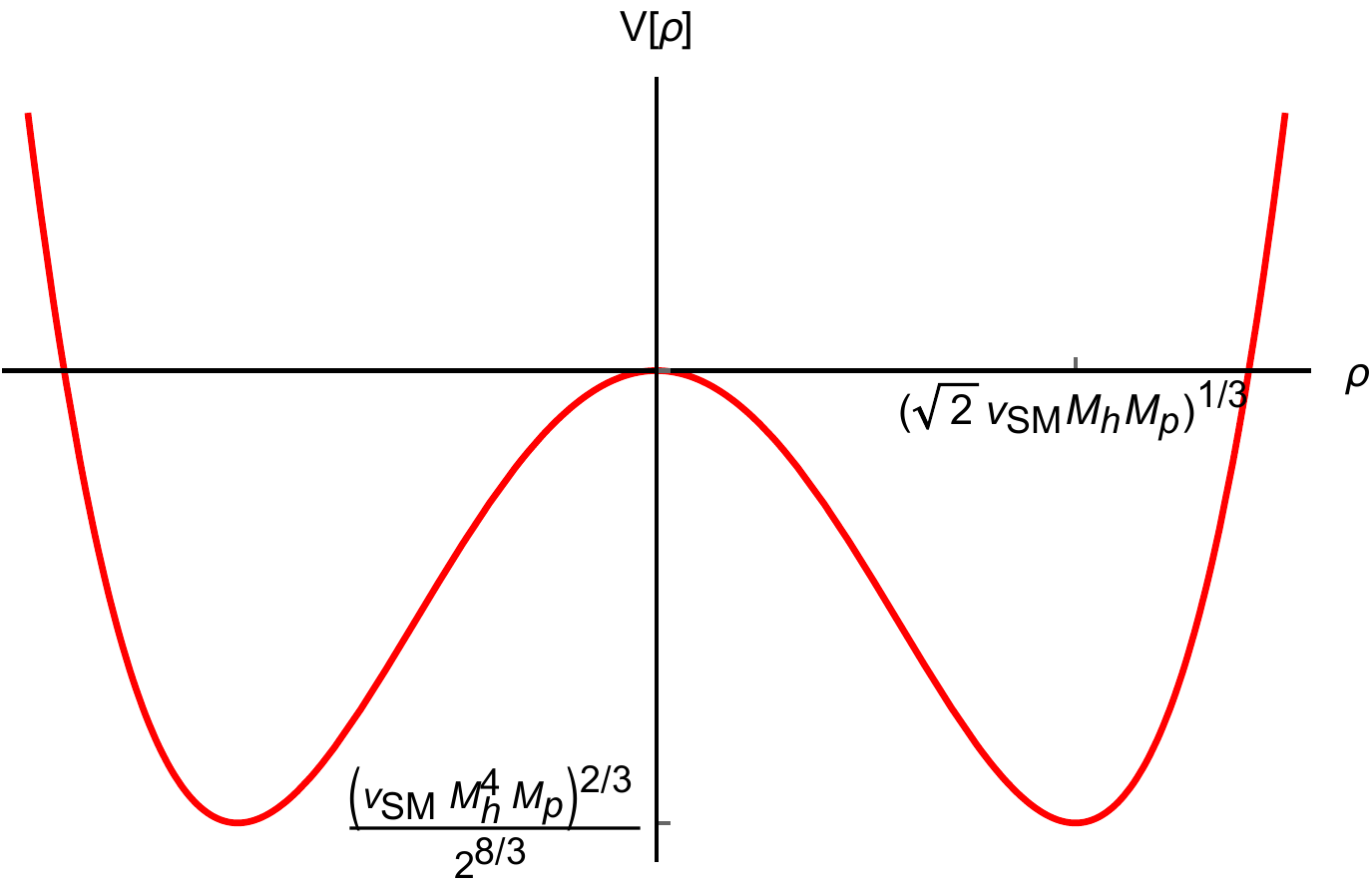}
     \caption{The sketch of potential defined from the EOM in Eq.~\eqref{VHiggss}, where $\phi(x)=\rho(x) e^{i\xi}/\sqrt{2}$}
     \label{fig:my_label}
 \end{figure}

To determine the value of $v$, we consider the dimension-6 operator contributing to the Higgs potential 
\begin{align}\label{29}
    \frac{\lambda\mu^2}{\Lambda^4}(\phi^*\phi)^3\;.
\end{align}
This operator provides the unitarity violation of amplitude at the energy $E>\Lambda^2/\sqrt{\lambda}\mu$. Therefore, the UV-cutoff can be specified at  $\Lambda_{\rm UV}\sim \Lambda^2/\sqrt{\lambda}\mu$.
If we assume that this model is an effective theory below the the Planck scale, it is reasonable to take the $\Lambda_{\rm UV }$ to be the reduced Planck mass ($M_p=2.44\times 10^{18}$ GeV). We have
\begin{align}\label{30}
   M_p^2=\frac{\Lambda^4}{\lambda\mu^2}.
\end{align}
Substituting  Eq.~\eqref{parameter1} into Eq.~\eqref{30} then applying the limit $v\gg v_{\rm SM}$, the vev is given by
\begin{align}\label{vevfit3}
  v\simeq (\sqrt{2}v_{\rm SM}M_h M_p)^{1/3}\sim 10^{7}\;\text{GeV}.
\end{align}
 We shall discuss the implication of the modified value of the Higgs vev in  the next section. Here, the structure of the Higgs potential in this model is different from the standard theory because the dimensionless coupling constant in our model is $\lambda= M_h^2/2v^2\sim 10^{-11}$ while in the standard theory is $\lambda= M_h^2/2v_{\rm SM}^2\sim 10^{-1}$.  The local minimum point of the Higgs potential, which is defined from the EOM, is shifted from the electroweak scale to a new scale at $10^7$ GeV, shown in Fig.~\ref{fig:my_label}
 However,  by substituting $\Lambda$ to  Eq.~\eqref{parameter1}, these parameters 
\begin{align}\label{26}
   & M_A=e v_{\rm SM}, \; y=\frac{M_f}{2v_\text{SM}},\;\lambda_3=\frac{3
   M_h^2}{v_{\text{SM}}},\; g_{hhA}=e^2 v_{\rm SM},
\end{align}
identically remain the same with the standard theory. The result of the Yukawa coupling is the same form with the SM and also agrees with the experimental observation \cite{sirunyan2019measurement,sirunyan2020measurement,aad2015evidence,sirunyan2018observation,aaboud2019cross,sirunyan2018evidence}. Therefore, the Higgs vev can be set into arbitrary high-energy scale without the sensitivity to these parameters in tree-level.

Then, we come to a major point of this research. We consider the UV cutoff energy of Higgs particle after SSB.  In the case of multiple values of the cutoff scale of higher dimension operators,
\begin{align}
    \frac{h^5}{\Lambda_5}+\frac{h^6}{\Lambda_6^2}+\mathcal{O}(h^7),
\end{align}
the $\Lambda_{\rm UV}$ would be determined from the minimal value of $\Lambda_i$ \cite{Manohar:1983md,Gavela:2016bzc,PhysRevD.79.081302,Burgess2010OnHI,Bezrukov:2010jz} to preserve the whole amplitude in the traditional perturbation theory.  We find that  $\Lambda_5\gg\Lambda_6$ so  $\Lambda_{\rm UV}$ can be effectively defined from $\Lambda_6$. Substituting $\Lambda$ in Eq.~\eqref{fitLambda} into $\Lambda_6$ in Eq.~\eqref{parameter1} and considering the limit $v\gg v_{\rm SM}$, we find that the UV-cutoff of Higgs boson after SSB is given by 
\begin{align}\label{lambdaUVbroken}
    \Lambda_{\rm UV}\simeq \frac{24\pi}{\sqrt{37}}\left(\frac{v_{\rm SM}}{M_h}\right)v_{\rm SM}+\mathcal{O}(v^{-1})\sim 10^3\text{GeV},
\end{align}
The UV-cutoff in the broken phase depends solely on the Fermi-coupling constant. Therefore, it is insensitive to the large scale of the Higgs vev.  Here, from Eq.~\eqref{lambdaUVbroken}, the ratio of the physical Higgs mass and the UV-cutoff in the broken phase is naturally of order unity
\begin{align}
    \frac{M_h^2}{\Lambda_{\rm UV}^2}\sim \mathcal{O}(1).
\end{align}
This result respects to the notion of naturalness. Obviously, the radiative correction from the SM-particles to the Higgs mass, Eq.~\eqref{1}, does not require the fine-tuning. 

In conclusion, the UV-cutoff before the SSB at the Planck scale \eqref{30} or arbitrary high energy scale can be tuned down exponentially to the TeV scale through the exponential constant factor, see Eq.~\eqref{beta}. 
\section{Discussions}
Firstly, we shall discuss on the deviation of the trilinear coupling ($\lambda_3$) and quartic Higgs self-coupling ($\lambda_4$) from the standard theory. We firstly consider the parameter $\lambda_3$ defined in Eq.~\eqref{26},
\begin{align*}
    \lambda_3=\frac{3M_h^2}{v_{\rm SM}^2}.
\end{align*}
This parameter is not altered from the SM prediction. This result is still available with the observed value at ILC and HL-LHC with the 68$\%$ confident level \cite{DiVita:2017vrr}. However, the observed value of this parameter does not come to the conclusion at the HL-LHC \cite{Observation, Observation2, Observation3, Observation4, Observation5, Observation6, Observation7, Observation8, Observation9, Observation10, Observation11}. Then, we consider the parameter $\lambda_4$. By substituting $\Lambda$ in Eq.~\eqref{fitLambda} into $\lambda_4$ in Eq.~\eqref{parameter1}, we obtain 
\begin{align}
   \lambda_4=\left(1-\frac{8 \left(v^2-v_{\text{SM}}^2\right) }{3
   v_{\text{SM}}^2}\log \left(\frac{v^2}{v^2-v_{\text{SM}}^2}\right)\right)\lambda_{\rm 4,SM},
\end{align}
where  $\lambda_{\rm 4,SM}=3M_h^2/v_{\rm SM}^2$ as the standard parameter from the Higgs mechanism in U(1)-gauge symmetry breaking. In the limit $v\gg v_{\rm SM}$, this coupling constant can be approximated as
\begin{align}\label{lambda4A}
 \lambda_4=\left(-\frac{5}{3}+O(\frac{v_{\rm SM}^2}{v^{2}})\right)\lambda_{\rm 4,SM}.
\end{align}
This result follows the unitarity bound of $hh\to hh$ $, | \lambda_4/\lambda_{\rm 4, SM}|<68$ \cite{Maxi-sizing, Probing-quartic}. The large scale of vev does not violate the perturbativity of scattering amplitude. Here, the result \eqref{lambda4A} will play a major role to distinguish our model from the other Higgs models, such as the Nambu-Goldstone Higgs boson \cite{Kaplan:1983fs,Kaplan:1983sm}, the Coleman-Weinberg Higgs boson \cite{PhysRevD.89.073003,Helmboldt:2016mpi,Hashino:2015nxa}, and the Tadpole-induced Higgs boson \cite{Galloway:2013dma,Chang:2014ida}, which predict various values of $\lambda_4/\lambda_{\rm 4, SM}$ \cite{PhysRevD.101.075023}.  However, the measurement of the quartic Higgs self-coupling is known to be difficult, even at a future 100 TeV hardron collider \cite{PhysRevD.72.053008}, and therefore the value of $\lambda_4$ is still far from the conclusion. 

Secondly, we discuss on the UV-sensitivity of the Higgs mass due to the (heavy) particle. The ratio of the one-loop correction from particle with mass $M$ and the physical Higgs mass ($\Delta\mu^2/M_h^2$) is given by
\begin{align}\label{41}
    &\frac{M^2 }{16 \pi ^2 M_h^2 v_{\text{SM}}^2}\left(M^2 \log \left(\frac{1}{2}+\frac{\Lambda _{\text{UV}}^2}{2M^2}\right)+M^2-\Lambda
   _{\text{UV}}^2\right),
\end{align}
where $\Lambda_{\rm UV}\sim v^2_{\rm SM}/M_h$. If the contribution comes from the SM particle such as top quark with mass $M=175$ GeV, the size of radiative correction is around $3\%$ of the size of physical Higgs mass. The bare mass parameter of Higgs and the physical Higgs mass are roughly close.  Thus, the radiative correction from SM particles does not violate the notion of naturalness.   Here, if the mass of the heavy particle is far beyond the electroweak scale $M\gg v_{\rm SM}$,  the remaining term is proportional to the heavy particle mass
\begin{align}
    \frac{\Delta\mu^2}{M_h^2}\simeq\frac{1}{16 \pi ^2}\left(\frac{M^2}{M_h^2}\right)\left(\frac{M^2}{v_{\rm SM}^2}\right).
\end{align}
Here, if $M\gg M_h$, there exists the fine-tuning problem in the bare Higgs mass. The notion of naturalness can preserved under the condition: $M^2/M_h^2\sim \mathcal{O}(1)$ or $M^2/v_{\rm SM}^2\sim \mathcal{O}(1)$.   However, if heavy fermions, coupling to the Higgs boson, actually exist in the experimental observation, the radiative correction no longer satisfies the notion of naturalness. The small value of the cutoff does not improvise anything in this situation. For example, if there is the existence of the particle with $M\sim M_p$, the size of the loop correction is around $\Delta\mu^2/M_h^2\sim 10^{30}$. Then, there is a huge amount of fine-tuning in the bare parameter of Higgs mass and, of course, the Higgs mass could get the UV-sensitivity from the high energy degree of freedom. The framework of our model  has to be replaced with more fundamental theory. For example,  supersymmetric theory or little Higgs model, the mass of scalar field can be protected from the large radiative correction.

Thirdly, we discuss on the application of our framework to provide the particularly small value of the parameter $\theta$ in the strong CP problem \cite{PhysRevLett.38.1440,Nelson:1983zb,Peccei:1988ci,Peccei:2006as}. It is well known that this problem requires the CP-breaking term into the QCD Lagrangian as
\begin{align}
    \frac{\theta}{16\pi^2} G^a_{\mu\nu} \tilde{G}^{\mu\nu a},
\end{align}
where $G^a_{\mu\nu}$ and $\tilde{G}^{\mu\nu a}$ are the QCD field strength and its dual. The current experimental data concerning the neutron electric dipole model introduces the upper limit of $\theta$, namely $\theta\lesssim 10^{-10}$ \cite{PhysRevD.86.036002,Pendlebury:2015lrz,Wurm:2019yfj}. In principle, this parameter can take any value between 0 to $2\pi$ but the nature selects the particularly small value very close to zero.  The Dirac naturalness requires the dimensionless parameter in order unity so this situation violates the naturalness expectation by the ten orders of magnitude. In this discussion, we study the effect of the addition of the operators, inside and outside, to the multiplicative Lagrangian. We try to explain the anomalous smallness of the $\theta$ term though the scheme. Here, we introduce the term $G^a_{\mu\nu} \tilde{G}^{\mu\nu a}/16\pi^2$ with the coefficient in the order unity, inside and outside, to the multiplicative Lagrangian as
\begin{align}\label{LLL}
    &\mathcal{L}=-\frac{1}{4} F_{\mu\nu} F^{\mu\nu}+D_\mu\phi^*D^\mu\phi-V+\frac{1}{16\pi^2} G^a_{\mu\nu} \tilde{G}^{\mu\nu a}\nonumber
    \\
    &-\left(-\Lambda^4+D_\mu\phi^*D^\mu\phi+\frac{1}{16\pi^2} G^a_{\mu\nu} \tilde{G}^{\mu\nu a}\right)e^{\frac{V}{\Lambda^4}}\nonumber
    \\
    &+i\bar{\psi}_L \slashed{D} \psi_L+i\bar{\psi}_R\slashed{D}\psi_R-\lambda_f\left(\phi\bar{\psi}_L\psi_R+\phi^*\bar{\psi}_R\psi_L\right).
\end{align}
When the SSB is taking place, the Higgs field can be expanded around the background value $\bar{\phi}$ as $\phi\to\bar{\phi}+h$, where  $h$ is the quantum fluctuation. The CP-breaking term depends on the background value of the Higgs field given by
\begin{align}
     \frac{1}{16\pi^2}\left(1-  e^{-\frac{\bar{\phi}^2 M_h^2}{8 \Lambda ^4  }}\right)G^a_{\mu\nu} \tilde{G}^{\mu\nu a}.
\end{align}
When $\bar{\phi}=0$, the value of the parameter $\theta$ is exact zero. Consequently, the CP-symmetry is restored at the origin. This process satisfies the notion of the technical naturalness. Then, when $\bar{\phi}=v$, $\theta$ becomes
\begin{align}\label{CPVourmodel}
  \theta= 1-  e^{-\frac{v^2 M_h^2}{8 \Lambda ^4 }}.
\end{align}
  Substituting Eq.~\eqref{fitLambda} into Eq.~\eqref{CPVourmodel} and applying $v\sim 10^7$ GeV from Eq.~\eqref{vevfit3}, the $\theta$ is rewritten as
\begin{align}\label{CPtheta}
    \theta=\frac{v_{\rm SM}^2}{v^2}\sim 10^{-10}.
\end{align}
This model can naturally give $\theta$ near the upper bound limit of the observed value without the introduction of a new hypothetical particle: axion \cite{PhysRevLett.38.1440,Nelson:1983zb,Peccei:1988ci,Peccei:2006as}. In conclusion, the addition of terms, with the same coefficient, inside and outside the multiplicative Lagrangian leads to the small coefficient after the SSB takes place. The thing is that the Lagrangian Eq.~\eqref{LLL} is our specific choice to explain the anomalous smallness of $\theta$ term in the strong CP problem.  However, the other choice of the Lagrangian, containing 
\begin{align}\label{57}
    -\frac{1}{4\pi} F_{\mu\nu}F^{\mu\nu}e^{\frac{V}{\Lambda^4}},\;\text{or} \; \lambda_f(\phi\bar{\psi}_L\psi_R+h.c.)e^{\frac{V}{\Lambda}},
\end{align}
without any symmetry violation, can be considered. We notice that these terms do not contribute to the $\theta$ parameter and, then, we shall ignore them.  Futhermore, we are going to show that if the Yukawa coupling, the third term in Eq.~\eqref{57}, is applied in this scheme, we also obtain some interesting context for the fermion mass parameter.

Fourthly, we discuss on the application of our framework to provide the fermion with anomalously small mass.  Suppose there is another type of fermion field $\chi$ with the Yukawa coupling  written inside and outside the multiplicative Lagrangian as
\begin{align}\label{Lnu}
   &\mathcal{L}=-\frac{1}{4} F_{\mu\nu} F^{\mu\nu}+D_\mu\phi^*D^\mu\phi-V\nonumber
    \\
    &+\alpha\left(\epsilon\Lambda^4+D_\mu\phi^*D^\mu\phi-\lambda_\chi\left(\phi\bar{\chi}_L\chi_R+\phi^*\bar{\chi}_R\chi_L\right)\right)e^{-\frac{V}{\epsilon\Lambda^4}}\nonumber
    \\
    &+i\bar{\psi}_L \slashed{D} \psi_L+i\bar{\psi}_R\slashed{D}\psi_R-\lambda_f\left(\phi\bar{\psi}_L\psi_R+\phi^*\bar{\psi}_R\psi_L\right)\nonumber
     \\
    &+i\bar{\chi}_L \slashed{D} \chi_L+i\bar{\chi}_R\slashed{D}\chi_R-\lambda_\chi\left(\phi\bar{\chi}_L\chi_R+\phi^*\bar{\chi}_R\chi_L\right).
\end{align}
After the SSB is taking place, mass of the $\chi$ field is given by
\begin{align}\label{massneutrino}
    m_\chi=\frac{v \lambda _{\chi } }{\sqrt{2}}\left(1-  e^{-\frac{v^2 M_h^2}{8 \Lambda ^4 }}\right).
\end{align}
Substituting Eq.~\eqref{fitLambda} into Eq.~\eqref{massneutrino} and considering $v\gg v_{\rm SM}$, the mass of particle $\chi$ is suppressed by the large scale of the Higgs vev as
\begin{align}
    m_\chi\approx\frac{\lambda _{\chi } v_{\text{SM}}^2}{\sqrt{2} v},
\end{align}
which is very small comparing with the mass of $\psi$ ($M_f$)
\begin{align}
    \frac{m_\chi}{M_f}=\left(\frac{\lambda_\chi}{\lambda_f}\right)\left(\frac{v_{\rm SM}^2}{v^2}\right)\sim 10^{-10}\left(\frac{\lambda_\chi}{\lambda_f}\right).
\end{align}
If we suppose that $\psi$ is an electron field with mass $M_f=0.5$ MeV and $\lambda_\chi/\lambda_f$ is of order unity (between $1-10^3$), the value of $m_\chi$ is  around
\begin{align}
    m_\chi\sim 0.0001- 0.1 \quad \text{eV}.
\end{align}
The upper bounded of $m_\chi$ implicitly relates to the observed  mass scale of the electron neutrino \cite{PhysRevLett.123.081301,KATRIN:2019yun,Schluter:2020gdr} with $\lambda_\chi\sim 10^{3}$. In our humble opinion, this toy mechanism could possibly be used to explain the neutrino mass  according to the Dirac Naturalness $\lambda_f/\lambda_\chi\sim \mathcal{O}(1)$.

Fifthly, we discuss the cosmological constant problem in this model. After SSB happens, the cosmological constant term generated from the Higgs potential is
\begin{align}
    \rho_{\rm Higgs}=\frac{1}{8} M_h^2 \left(\frac{v_{\text{SM}}^2-v^2}{\log \left(\frac{v^2}{v^2-v_{\text{SM}}^2}\right)}-v^2\right).
\end{align}
In the limit $v\gg v_{\rm SM}$, we have
\begin{align}
    \rho_{\rm Higgs}=& -\frac{v^4 M_h^2}{8 v_{\text{SM}}^2}+O\left(v^{2}\right).
\end{align}
 If  $v=10^7$ GeV, we obtain $\rho_{\rm Higgs}\sim 10^{27}$ (GeV)$^4$. This is obviously larger than the observed cosmological constant,  where the observed value is around $\rho_{\rm vac}\sim 10^{-47}$ (GeV)$^4$ \cite{aghanim2020planck}, about 74 orders of magnitude.  The easiest possible way to improvise this situation is an "ad hoc" constant term $\rho_X$ into the Lagrangian \eqref{Ltoy1}, as the same trick in Higgs mechanism \cite{RevModPhys.61.1},
\begin{align}
    \mathcal{L}'=\mathcal{L}+\rho_X.
\end{align}
Here,  we can set $\rho_X=\rho_{\rm Higgs}-\rho_{\rm vac}$ to cancel all the tremendous contributions. However, this process leads to a huge fine-tuning of the parameter $\rho_X$ and the infinitesimal value of the cosmological constant is not predictable. Therefore, the linear combination between the standard Lagrangian and the multiplicative Lagrangian could not provide a resolution to the cosmological constant. 

Lastly, we shall discuss on the energy density of the theory defined from the Hamiltonian density. From the Lagrangian \eqref{LinearCL2}, the Hamiltonian density ($\mathcal{H}$), defined as the energy density of the theory, is given by
\begin{align}\label{energyoftheory}
    \mathcal{H}=\left(1-e^{\frac{V}{\Lambda^4}}\right)\left(\dot{\phi}^*\dot{\phi}+\nabla\phi^*\cdot\nabla\phi \right)+V_{\rm eff},
\end{align}
where 
\begin{align}\label{veff}
    V_{\rm eff}=V-\Lambda^4e^{\frac{V}{\Lambda^4}}
\end{align}
will be called the effective potential. This potential has five extremum points at $|\rho|=0,\pm v,\pm\sqrt{2} v$, where $\phi(x)=\rho(x)\exp(i \xi(x))/\sqrt{2}$.  The potential start to decrease down to $-\infty$ when $|\rho|>\sqrt{2}v$. Therefore, the effective potential is unbounded from below. Hence, the vacuum solution of field from the EOM, $v=\mu/\sqrt{\lambda}$, is not the true vacuum. This situation leads to the vacuum instability problem since the Higgs particle can possibly tunnel from the meta-stable state at $v$ to unbounded negative energy state. This risky situation can be improved by adding the extra potential $V$ term inside the multiplicative Lagrangian \eqref{Ltoy1} as
\begin{align}\label{newL}
    \mathcal{L}'=\mathcal{L}+V e^{\frac{V}{\Lambda^4}}.
\end{align}
This new Lagrangian \eqref{newL} still processes the $U(1)$ symmetry. Here, the effective potential \eqref{veff} becomes
\begin{align}\label{veff2}
    V_{\rm eff}=V-\left(\Lambda^4+V \right)e^{\frac{V}{\Lambda^4}}.
\end{align}
Now, we have $V$ inside and outside the multiplicative Lagrangian, motivated from Eq.~\eqref{LLL} and Eq.~\eqref{Lnu}. Of course, a new EOM, obtained from Eq.~\eqref{newL}, will come with extra interaction terms produced by $V\exp{V/\Lambda^4}$. However, the change of the EOM does not matter at this level since only the Lagrangian \eqref{newL} will be used to analyse the physical parameters in the model. When the kinetic energy is zero, the remaining terms in the energy density \eqref{energyoftheory} is the potential \eqref{veff2}.  The tree level Higgs vev from Eq.~\eqref{veff2} is $v^2=\mu^2/\lambda$. After all parameters in the model are fit with the Fermi coupling constant, we find that the Higgs vev $v$ is still free from the parameter $G_F$. However, in the limit $v\gg v_{\rm SM}$, the sign of the parameters $\mu^2$, $\lambda$, and $\Lambda^4$ turns to be negative. If $v\simeq 10^7$ GeV, we have $\mu^2\simeq -(2.2\times 10^{-3})^2$ GeV$^2$, $\lambda\simeq -4.7\times 10^{-20}$, and $\Lambda^4\simeq-( 2.1\times 10^4)^4$ GeV$^4$. To converse the negative sign to be a positive of the parameter $\Lambda^4$, the potential has to be re-defined as
\begin{align}
   V\to \tilde{V}=\tilde{\mu}^2 \phi^*\phi-\tilde{\lambda} (\phi^*\phi)^2,
\end{align}
where $\tilde{\mu}^2=-\mu^2$ and $\tilde{\lambda}=-\lambda$.
 The full effective potential is rewritten as
\begin{align}\label{veff3}
    V_{\rm eff}=\tilde{V}+\tilde{\Lambda}^4e^{-\frac{\tilde{V}}{\tilde{\Lambda}^4}}-\tilde{V} e^{-\frac{\tilde{V}}{\tilde{\Lambda}^4}},
\end{align}
where $\tilde{\Lambda}^4=-\Lambda^4$.
Now, the value of parameters $\tilde{\mu}^2$, $\tilde{\lambda}$, and $\tilde{\Lambda}^4$, which their expressions are shown in appendix, are all positive. When $\phi\to\pm\infty$, the new effective potential is dominated by the exponentially growth term $+\tilde{\lambda}(\phi^*\phi)^2\exp({\tilde{\lambda} (\phi^*\phi)^2/\tilde{\Lambda}^4})$. So, this potential is obviously bounded from below. Therefore, no vacuum instability exists in the tree-level analysis. We then discuss the major parameters in the model under the condition $v\sim 10^7$ GeV. Firstly, the UV-cutoff of Higgs field $\Lambda_6\sim 10^3$ GeV is still applicable and is still independent on $v$. The parameter $\lambda_3$ is in the same form but the parameter $\lambda_4$ is modified as $\lambda_4/\lambda_{4,\text{SM}}\simeq 6$, which is still allowed with the unitarity and the observation at the HL-LHC. A last important point,  our analysis result on the strong CP problem and the small fermion mass is still intact. Finally, the cosmological constant term is also modified as $\rho_{\rm Higgs}\simeq M_h^2 v^2/8$, which is still in the realm of the cosmological constant problem. Up to this point, the multiplicative Lagrangian \eqref{LinearCL2}, motivated from the inverse calculus of variation in our major analysis, can lead to the various predictions of the energy scale beyond SM. But, it is an unsatisfied Lagrangian due to the vacuum instability. However, the addition of $V$ inside and outside the multiplicative Lagrangian can fix the vacuum instability problem and the results of the rest are not spoiled.


\section{Conclusions}
We apply the idea of the non-uniqueness principle of the Lagrangian to the toy model of the electroweak phase transition. The multiplicative Lagrangian, respecting to the approximated Klein-Gordon equation, is promoted to conserve the autonomy of physical scale in the scalar field.  We find that the Higgs vev is free from the Fermi-coupling constant and can be tuned to be arbitrary high energy scale. However, if our model \eqref{Ltoy1} is an effective theory below the Planck scale, it possibly seems to give the prediction of the size of Higgs mass and the TeV cutoff scale in the broken phase. Lastly, this framework would presumably provide a preliminary step to alternatively explain the strong CP-problem and the neutrino mass mechanism. 


\begin{acknowledgements}
We acknowledge the support from the Petchra Prajomklao Ph.D. Research Scholarship from King Mongkut’s University of Technology Thonburi (KMUTT).
\end{acknowledgements}

\appendix
\section{}
\begin{align}
    &\tilde{\Lambda} ^4=\frac{\mu ^2 v^2}{4 \log \left(\frac{v^2}{v^2-v_{\text{SM}}^2}\right)}
    \\
    &\tilde{\mu} ^2=\frac{M_h^2 v_{\text{SM}}^2}{2 v^2}+O\left(v^{-4}\right)
    \\
    &\tilde{\lambda }=\frac{M_h^2 v_{\text{SM}}^2}{2 v^4}+O\left(v^{-6}\right)
    \\
   & \lambda _3=\frac{3 M_h^2}{v_{\text{SM}}}
    \\
    &\lambda _4=\frac{19 M_h^2}{v_{\text{SM}}^2}+O\left(v^{-2}\right)
    \\
   & \Lambda _5=\frac{8 \pi  v_{\text{SM}}^3}{3 M_h^2}+O\left(v^{-2}\right)
    \\
    &\Lambda_6=\frac{24 \pi  v_{\text{SM}}^2}{\sqrt{113} M_h}+O\left(v^{-2}\right)
    \\
    &\rho _{\text{Higgs}}=\frac{1}{8} v^2 M_h^2+\frac{3}{16} M_h^2 v_{\text{SM}}^2+O\left(v^{-2}\right)
\end{align}

\bibliographystyle{apsrev4-1}
\bibliography{mybib.bib}
\end{document}